\documentstyle[12pt]{article}
\newcommand{\nc}{\newcommand}
\nc{\renc}{\renewcommand}

\nc{\half}{{\textstyle{1\over2}}}
\nc{\etal}{\mbox{\it et al. }}
\nc{\ie}{{\it i.e.}}
\nc{\eg}{{\it e.g.}}

\renc{\thefootnote}{\arabic{footnote}}
\nc{\capt}[1]{{\bf Figure.} {\small\sl #1}}

\nc{\eqs}[2]{\mbox{Eqs.~(\ref{#1},\,\ref{#2})}}
\nc{\eq}[1]{\mbox{Eq.~(\ref{#1})}}

\nc{\figs}[2]{\mbox{Figs.~(\ref{#1},\,\ref{#2})}}
\nc{\fig}[1]{\mbox{Fig~.(\ref{#1})}}

\nc{\tag}[1]{\label{#1} \marginpar{{\footnotesize #1}}}
\nc{\mtag}[1]{\label{#1} \mbox{\marginpar{{\footnotesize #1}}}}
\renc{\baselinestretch}{1.5}
\newlength{\overeqskip}
\newlength{\undereqskip}
\nc{\be}[1]{\begin{equation} \mbox{$\label{#1}$}}
\nc{\bea}[1]{\begin{eqnarray} \mbox{$\label{#1}$}}
\nc{\Section}[2]{\section{#2}\label{#1}}
\nc{\Bibitem}[1]{\bibitem{#1}}
\nc{\Label}[1]{\label{#1}}

\nc{\eea}{\vspace{\undereqskip}\end{eqnarray}}
\nc{\ee}{\vspace{\undereqskip}\end{equation}}
\nc{\bdm}{\begin{displaymath}}
\nc{\edm}{\end{displaymath}}
\nc{\dpsty}{\displaystyle}
\nc{\bc}{\begin{center}}
\nc{\ec}{\end{center}}
\nc{\ba}{\begin{array}}
\nc{\ea}{\end{array}}
\nc{\bab}{\begin{abstract}}
\nc{\eab}{\end{abstract}}
\nc{\btab}{\begin{tabular}}
\nc{\etab}{\end{tabular}}
\nc{\bit}{\begin{itemize}}
\nc{\eit}{\end{itemize}}
\nc{\ben}{\begin{enumerate}}
\nc{\een}{\end{enumerate}}
\nc{\bfig}{\begin{figure}}
\nc{\efig}{\end{figure}}
\nc{\arreq}{&\!=\!&}
\nc{\arrmi}{&\!-\!&}
\nc{\arrpl}{&\!+\!&}
\nc{\arrap}{&\!\!\!\approx\!\!\!&}
\nc{\non}{\nonumber\\*}
\nc{\align}{\!\!\!\!\!\!\!\!&&}

\def\lsim{\; \raise0.3ex\hbox{$<$\kern-0.75em
      \raise-1.1ex\hbox{$\sim$}}\; }
\def\gsim{\; \raise0.3ex\hbox{$>$\kern-0.75em
      \raise-1.1ex\hbox{$\sim$}}\; }
\nc{\DOT}{\hspace{-0.08in}{\bf .}\hspace{0.1in}}
\nc{\Laada}{\hbox {$\sqcap$ \kern -1em $\sqcup$}}
\nc\loota{{\scriptstyle\sqcap\kern-0.55em\hbox{$\scriptstyle\sqcup$}}}
\nc\Loota{{\sqcap\kern-0.65em\hbox{$\sqcup$}}}
\nc\laada{\Loota}
\nc{\qed}{\hskip 3em \hbox{\BOX} \vskip 2ex}

\nc{\real}{{\rm I \! R}}
\nc{\Z}{{\sf Z \!\!\! Z}}
\nc{\complex}{{\rm C\!\!\! {\sf I}\,\,}}
\def\bigid{\leavevmode\hbox{\small1\kern-3.8pt\normalsize1}}
\def\id{\leavevmode\hbox{\small1\kern-3.3pt\normalsize1}}
\nc{\slask}{\!\!\!/}
\nc{\bis}{{\prime\prime}}
\nc{\pa}{\partial}
\nc{\na}{\nabla}
\nc{\ra}{\rangle}
\nc{\la}{\langle}
\nc{\goto}{\rightarrow}
\nc{\swap}{\leftrightarrow}

\nc{\EE}[1]{ \mbox{$\cdot10^{#1}$} }
\nc{\abs}[1]{\left|#1\right|}
\nc{\at}[2]{\left.#1\right|_{#2}}
\nc{\norm}[1]{\|#1\|}
\nc{\abscut}[2]{\Abs{#1}_{\scriptscriptstyle#2}}
\nc{\vek}[1]{{\rm\bf #1}}
\nc{\integral}[2]{\int\limits_{#1}^{#2}}
\nc{\inv}[1]{\frac{1}{#1}}
\nc{\dd}[2]{{{\partial #1}\over{\partial #2}}}
\nc{\ddd}[2]{{{{\partial}^2 #1}\over{\partial {#2}^2}}}
\nc{\dddd}[3]{{{{\partial}^2 #1}\over
	{\partial #2 \partial #3}}}
\nc{\dder}[2]{{{d #1}\over{d #2}}}
\nc{\ddder}[2]{{{d^2 #1}\over{d {#2}^2}}}
\nc{\dddder}[3]{{d^2 #1}\over
	{d #2 d #3}}
\nc{\dx}[1]{d\,^{#1}x}
\nc{\dy}[1]{d\,^{#1}y}
\nc{\dz}[1]{d\,^{#1}z}
\nc{\dl}[1]{\frac{d\,^{#1}l}{(2\pi)^{#1}}}
\nc{\dk}[1]{\frac{d\,^{#1}k}{(2\pi)^{#1}}}
\nc{\dq}[1]{\frac{d\,^{#1}q}{(2\pi)^{#1}}}

\nc{\cc}{\mbox{$c.c.$ }}
\nc{\hc}{\mbox{$h.c.$ }}
\nc{\cf}{cf.\ }
\nc{\erfc}{{\rm erfc}}
\nc{\Tr}{{\rm Tr\,}}
\nc{\tr}{{\rm tr\,}}
\nc{\pol}{{\rm pol}}
\nc{\sign}{{\rm sign}}
\nc{\bfT}{{\bf T }}

\def\GeV{{\rm\ GeV}}

\nc{\cA}{{\cal A}}
\nc{\cB}{{\cal B}}
\nc{\cD}{{\cal D}}
\nc{\cE}{{\cal E}}
\nc{\cG}{{\cal G}}
\nc{\cH}{{\cal H}}
\nc{\cL}{{\cal L}}
\nc{\cO}{{\cal O}}
\nc{\cT}{{\cal T}}
\nc{\cN}{{\cal N}}
\nc{\rvac}[1]{|{\cal O}#1\rangle}
\nc{\lvac}[1]{\langle{\cal O}#1|}
\nc{\rvacb}[1]{|{\cal O}_\beta #1\rangle}
\nc{\lvacb}[1]{\langle{\cal O}_\beta #1 |}
\nc{\bb}{\bar{\beta}}
\nc{\bt}{\tilde{\beta}}
\nc{\ctH}{\tilde{\cal H}}
\nc{\chH}{\hat{\cal H}}
\nc{\1}{\aa}
\nc{\2}{\"{a}}
\nc{\3}{\"{o}}
\nc{\4}{\AA}
\nc{\5}{\"{A}}
\nc{\6}{\"{O}}
\nc{\al}{\alpha}
\nc{\g}{\gamma}
\nc{\Del}{\Delta}
\nc{\e}{\epsilon}
\nc{\eps}{\epsilon}
\nc{\lam}{\lambda}
\nc{\om}{\omega}
\nc{\Om}{\Omega}
\nc{\ve}{\varepsilon}
\nc{\mn}{{\mu\nu}}
\nc{\k}{\kappa}
\nc{\vp}{\varphi}

\nc{\advp}[3]{{\it  Adv.\ in\ Phys.\ }{{\bf #1} {(#2)} {#3}}}
\nc{\annp}[3]{{\it  Ann.\ Phys.\ (N.Y.)\ }{{\bf #1} {(#2)} {#3}}}
\nc{\apl}[3]{{\it  Appl. Phys. Lett. }{{\bf #1} {(#2)} {#3}}}
\nc{\apj}[3]{{\it  Ap.\ J.\ }{{\bf #1} {(#2)} {#3}}}
\nc{\apjl}[3]{{\it  Ap.\ J.\ Lett.\ }{{\bf #1} {(#2)} {#3}}}
\nc{\app}[3]{{\it Astropart.\ Phys.\ }{{\bf #1} {(#2)} {#3}}}
\nc{\cmp}[3]{{\it  Comm.\ Math.\ Phys.\ }{{ \bf #1} {(#2)} {#3}}}
\nc{\cqg}[3]{{\it  Class.\ Quant.\ Grav.\ }{{\bf #1} {(#2)} {#3}}}
\nc{\epl}[3]{{\it  Europhys.\ Lett.\ }{{\bf #1} {(#2)} {#3}}}
\nc{\ijmp}[3]{{\it Int.\ J.\ Mod.\ Phys.\ }{{\bf #1} {(#2)} {#3}}}
\nc{\ijtp}[3]{{\it Int.\ J.\ Theor.\ Phys.\ }{{\bf #1} {(#2)} {#3}}}
\nc{\jmp}[3]{{\it  J.\ Math.\ Phys.\ }{{ \bf #1} {(#2)} {#3}}}
\nc{\jpa}[3]{{\it  J.\ Phys.\ A\ }{{\bf #1} {(#2)} {#3}}}
\nc{\jpc}[3]{{\it  J.\ Phys.\ C\ }{{\bf #1} {(#2)} {#3}}}
\nc{\jap}[3]{{\it J.\ Appl.\ Phys.\ }{{\bf #1} {(#2)} {#3}}}
\nc{\jpsj}[3]{{\it J.\ Phys.\ Soc.\ Japan\ }{{\bf #1} {(#2)} {#3}}}
\nc{\lmp}[3]{{\it Lett.\ Math.\ Phys.\ }{{\bf #1} {(#2)} {#3}}}
\nc{\mpl}[3]{{\it  Mod.\ Phys.\ Lett.\ }{{\bf #1} {(#2)} {#3}}}
\nc{\ncim}[3]{{\it  Nuov.\ Cim.\ }{{\bf #1} {(#2)} {#3}}}
\nc{\np}[3]{{\it  Nucl.\ Phys.\ }{{\bf #1} {(#2)} {#3}}}
\nc{\npps}[3]{{\it  Nucl.\ Phys.\ Proc.\ Suppl.\ }{{\bf #1} {(#2)} {#3}}}
\nc{\pr}[3]{{\it Phys.\ Rev.\ }{{\bf #1} {(#2)} {#3}}}
\nc{\pra}[3]{{\it  Phys.\ Rev.\ A\ }{{\bf #1} {(#2)} {#3}}}
\nc{\prb}[3]{{\it  Phys.\ Rev.\ B\ }{{{\bf #1} {(#2)} {#3}}}}
\nc{\prc}[3]{{\it  Phys.\ Rev.\ C\ }{{\bf #1} {(#2)} {#3}}}
\nc{\prd}[3]{{\it  Phys.\ Rev.\ D\ }{{\bf #1} {(#2)} {#3}}}
\nc{\prl}[3]{{\it Phys.\ Rev.\ Lett.\ }{{\bf #1} {(#2)} {#3}}}
\nc{\pl}[3]{{\it  Phys.\ Lett.\ }{{\bf #1} {(#2)} {#3}}}
\nc{\prep}[3]{{\it Phys.\ Rep.\ }{{\bf #1} {(#2)} {#3}}}
\nc{\prsl}[3]{{\it Proc.\ R.\ Soc.\ London\ }{{\bf #1} {(#2)} {#3}}}
\nc{\ptp}[3]{{\it  Prog.\ Theor.\ Phys.\ }{{\bf #1} {(#2)} {#3}}}
\nc{\ptps}[3]{{\it  Prog\ Theor.\ Phys.\ suppl.\ }{{\bf #1} {(#2)} {#3}}}
\nc{\physa}[3]{{\it  Physica\ A\ }{{\bf #1} {(#2)} {#3}}}
\nc{\physb}[3]{{\it  Physica\ B\ }{{\bf #1} {(#2)} {#3}}}
\nc{\phys}[3]{{\it Physica\ }{{\bf #1} {(#2)} {#3}}}
\nc{\rmp}[3]{{\it  Rev.\ Mod.\ Phys.\ }{{\bf #1} {(#2)} {#3}}}
\nc{\rpp}[3]{{\it Rep.\ Prog.\ Phys.\ }{{\bf #1} {(#2)} {#3}}}
\nc{\sjnp}[3]{{\it Sov.\ J.\ Nucl.\ Phys.\ }{{\bf #1} {(#2)} {#3}}}
\nc{\spjetp}[3]{{\it Sov.\ Phys.\ JETP\ }{{\bf #1} {(#2)} {#3}}}
\nc{\yf}[3]{{\it Yad.\ Fiz.\ }{{\bf #1} {(#2)} {#3}}}
\nc{\zetp}[3]{{\it Zh.\ Eksp.\ Teor.\ Fiz.\  }{{\bf #1}  {(#2)} {#3}}}
\nc{\zp}[3]{{\it Z.\ Phys.\ }{{\bf #1} {(#2)} {#3}}}
\nc{\ibid}[3]{{\sl ibid.\ }{{\bf #1} {#2} {#3}}}
\nc{\rf}[1]{(\ref{#1})}
\nc{\nn}{\nonumber \\*}
\nc{\bfB}{\bf{B}}
\nc{\bfv}{\bf{v}}
\nc{\bfx}{\bf{x}}
\nc{\bfy}{\bf{y}}
\nc{\vx}{\vec{x}}
\nc{\vy}{\vec{y}}
\nc{\oB}{\overline{B}}
\nc{\oI}{\overline{I}}
\nc{\oR}{\overline{R}}
\nc{\rar}{\rightarrow}
\nc{\ti}{\times}
\nc{\slsh}{\hskip-5pt/}
\nc{\sm}{Standard~Model~}
\nc{\MP}{M_{\rm Pl}}
\nc{\tp}{t_{\rm Pl}}
\nc{\ave}{\bar{E}}

\nc{\eff}{{\rm eff}}
\nc{\kk}{\vek{k}}
\nc{\pp}{{\rm p}}
\nc{\ga}{g_{a\gamma}}
\nc{\vv}{\\}
\nc{\eee}{{\bf E}}
\nc{\bbb}{{\bf B}}
\nc{\qcd}{T_{\rm QCD}}
\nc{\G}{\rm \ G}
\def\vec#1{{\bf #1}}
\def\vv{\vskip-2pt}

\def\ell{e^{c}LL}

\begin{document}
{\title{\vskip-2truecm{\hfill {{\small HIP-2000-18/TH\\
	\hfill \\
	}}\vskip 1truecm}
{\LARGE Flat direction condensate instabilities
in the MSSM}}
\vspace{-.2cm}
{\author{
{\sc Kari Enqvist$^{1}$}
\\
{\sl\small Physics Department, University of Helsinki, and Helsinki Institute 
of Physics}\\
{\sl\small  P.O. Box 9, FIN-00014 University of Helsinki, FINLAND}
\\
{\sc Asko Jokinen$^{2}$}
\\
{\sl\small Physics Department, University of Helsinki }\\
{\sl\small P.O. Box 9, FIN-00014 University of Helsinki, FINLAND}
\\
{\sc  John McDonald$^{3}$}\\
{\sl\small Department of Physics and Astronomy,
University of Glasgow  }\\
{\sl\small Glasgow G12 8QQ, SCOTLAND}
}
\maketitle
\begin{abstract}
\noindent
 Coherently oscillating scalar condensates formed along flat
directions of the MSSM scalar potential are unstable with respect to 
spatial perturbations if the potential is flatter than $\phi^2$,
resulting in the formation of non-topological solitons such as Q-balls.       
Using the renormalization group we calculate the corrections to the $\phi^2$
potential for a range of flat directions and show that unstable condensates
are a generic feature of the MSSM. Exceptions arise for an
experimentally testable range of stop and gluino masses when there are 
large admixtures of stops in the flat direction scalar.

\end{abstract}
\vfil
\footnoterule
{\small $^1$kari.enqvist@helsinki.fi;  $^2$asko.jokinen@helsinki.fi;
$^3$mcdonald@physics.gla.ac.uk}

\thispagestyle{empty}
\newpage
\setcounter{page}{1}

\section{Introduction}

        The scalar potential of the Minimal Supersymmetric Standard Model (MSSM)
 \cite{nilles} is a complicated function of 45 complex squark and slepton fields and 4
 complex Higgs fields. A natural feature of this potential is the existence of flat directions,
 corresponding to linear combinations of these fields such that there are no renormalizable
 contributions to the scalar potential along the flat direction beyond 
the soft SUSY breaking terms \cite{drt}. These flat directions may play a fundamental role
 in the 
cosmology of the MSSM, in particular as a natural source of the baryon asymmetry of the
 Universe via the Affleck-Dine mechanism \cite{ad}, in which a baryon asymmetry is
 induced in a coherently oscillating condensate of squarks and sleptons. It has recently been
 realized that the cosmology of flat directions of the MSSM and Affleck-Dine baryogenesis
 may be more complicated than previously thought \cite{k,ksad,ksdm,bbb1,bbb2}. If the flat
 direction scalar potential increases less rapidly than $\phi^{2}$, where $\phi$ is the scalar
 field along the flat direction ("Affleck-Dine (AD) scalar"), then once coherent oscillations of the
 field begin, 
the condensate will have a negative pressure making it unstable with respect to spatial
 perturbations \cite{ksad,bbb1}. These spatial perturbations, which generally arise during
 inflation as a result of quantum fluctuations of the AD scalar \cite{bbb1,bbb2,bbdyn}, will
 grow and go non-linear, fragmenting the condensate into lumps which eventually
 evolve into Q-balls of baryon number ("B-balls") \cite{ksad,bbb1,bbdyn}. 
The formation of Q-balls has recently been verified by lattice simulations \cite{lattice}.

The
 cosmological evolution of the resulting Q-balls depends on the 
form of SUSY breaking. Q-balls from the AD condensate were first discussed in the 
context of gauge-mediated SUSY breaking \cite{ksad,ksdm}, in which the flatness of the potential
 occurs above the mass of the messenger fields. It was later realized that Q-balls can also form in the
 more conventional gravity-mediated SUSY breaking models \cite{bbb1,bbb2}. In this case
 the potential 
 is likely to be flatter than $\phi^{2}$ as a result of radiative corrections. The resulting Q-balls are not stable but can be very long-lived, decaying after the electroweak phase 
transition has occured \cite{bbb2}. As a result Q-balls can protect the baryon asymmetry from the 
effect of lepton number violating interactions combined with sphaleron processes and, if they
 decay at a low enough temperature, can also act as a common source of baryons and dark
 matter neutralinos, so explaining the similarity of the number densities of baryons and dark matter particles when the 
 dark matter particles have masses of the order of $m_{W}$ \cite{bbb2,bbbdm}. They can
 also enhance the isocurvature density perturbations expected from AD baryogenesis in
 particular inflation models \cite{bbiso,bbiso2}. 

              The existence and cosmology of Q-balls in the MSSM
 with gravity-mediated SUSY breaking is dependent upon the details of the radiative correction to the flat direction scalar
 potential. The correction must be negative in sign relative to the SUSY breaking mass
 squared term in order for Q-balls to form, whilst the binding energy of the charges within the Q-ball
 and so the total charge and lifetime of the Q-ball depends upon the magnitude of the
 correction \cite{bbb2}, as does the fraction of the total baryon number initially trapped
 within the Q-balls \cite{bbdyn}. In this 
 letter we consider in detail the radiatively corrected scalar potential for a range of flat directions using
 the renormalization group. 

We will show that condensate collapse and Q-ball formation are
 almost a general feature of all the 
MSSM flat directions, with exceptions only for the cases of d=4
$H_uL$-direction and 
 directions with large admixtures of stop. The existence of 
instabilities along
the latter directions  depend on the mass of the stop in a
testable way, as will be discussed in the following.

F- and D-flat directions of the MSSM have been classified and listed in \cite{drt}.
For gravity-mediated SUSY breaking the scalar potential along a
flat direction has the form \cite{bbb1,bbb2}
\be{pot} U(\Phi) \approx m^{2}\left(1 +  
K \log\left( \frac{|\Phi|^{2}}{M^{2}}
 \right) \right) |\Phi|^{2} 
+ \frac{\lambda^{2}|\Phi|^{2(d-1)}
}{M_{*}^{2(d-3)}} + \left( \frac{A_{\lambda} 
\lambda \Phi^{d}}{d M_{*}^{d-3}} + h.c.\right)    ~,
\ee
where $m$ is the conventional gravity-mediated soft SUSY breaking scalar 
mass term
 ($m \approx 100 \GeV$), $K$ is a parameter which depends on the
flat direction, and
$d$ is the dimension of the non-renormalizable term in the superpotential
which first lifts the degeneracy of the flat direction; it is of the form $f=\lambda 
M_*^{4-d}\Phi_1\cdots \Phi_d$.
We assume that the natural scale of the 
non-renormalizable terms is $M_{*}$, where 
$M_{*} = M_{Pl}/\sqrt{8 \pi}$ is the supergravity mass scale. 

The equation of state for a field oscillating in a potential $\phi^\gamma$
reads as \cite{pressure} $p = (\frac{2\gamma}{\gamma+2}  - 1)\rho$
so that \eq{pot} gives rise to the equation of
state $p = \frac{K}{2} \rho$. With $K<0$ the pressure is negative and hence the AD 
condensate is unstable. The condensate fragments and eventually
the lumps will evolve dynamically into the state of lowest energy,  the Q-ball.
If $K>0$ the condensate is stable and no Q-balls will form. Although
a negative $K$ should be a generic feature of the MSSM, not all flat directions
will have negative $K$ for all the values of the MSSM parameters. Moreover,
as the actual value of $K$ dictates the dynamical evolution of the AD condensate
and its fragmentation, it is
of great interest to find out the precise value of $K$ for
a given flat direction and for a range of the MSSM parameter values.

$K$ can be computed from the RG equations, which to one loop have
the form
\be{rge}
{\partial m_i^2\over \partial t}=\sum_{g}a_{ig} m_g^2+
\sum_a h^2_a(\sum_j b_{ij}m_j^2+A^2)~,
\ee
where $a_{ig}$ and $b_{ij}$ are constants, $m_g$ is the gaugino mass, $A$ is
the A-term, $h_a$ the Yukawa coupling, and $t=\ln M_X/\mu$.
The full RG equations are listed in \cite{nilles} and we do not reproduce them 
here.
We assume unification at $t=0$ and
neglect all other Yukawa couplings except the top Yukawa $h_t(M_W)=1$
(for definiteness, we choose $tan\beta=1$). 
We shall use quantities scaled by $m$ and denote $m_g/m$ at $t=0$ by $\xi$. 
The potential
along the flat direction is then
characterized by the amount of stop mixture (where appropriate), the values
of $\xi$ and $A$, and in the special case of the d=4 $H_uL$ -direction,
on the $H_u H_d$ -mixing mass parameter $\mu_H$.

The mass of the AD scalar $\phi$ is the sum of the masses of the 
squark and slepton fields $\phi_i$ constituting
the flat direction, 
$
m_S^2 = \sum_{a} p_i^2 m_i^2~,
$
where $p_i$ is the projection of $\phi$ along $\phi_i$, and $\sum p_i^2=1$.
The parameter $K$ is then given simply by
\be{K1}
K={\partial m_S^2\over \partial t}\Big\vert_{t={\rm log}\mu}~.
\ee

To compute $K$, we have to choose the scale $\mu$. The appropriate scale
is given by the value of the AD field when it first begins to oscillate
at $H\approx m$. Let the value of the mean field be $\phi_0$; the value of $K$
at this scale then determines whether the condensate is unstable or not. 
We may compute $\phi_0$ from \eq{pot} by ignoring the radiative correction (i.e. setting
effectively $K=0$) and minimizing 
the U(1) symmetric part of $U$ (or neglecting the A-term in
\eq{pot}). One then finds
\be{phi}
\mu=|\phi_0| = \left[\frac{m^2 M_*^{2(d-3)}}{(d-1)\lambda^2}\right]^{\frac{1}{2(d-2)}}~,
\ee
where in what follows we assume for simplicity that $\lambda=1$.

\begin{center} {\bf Table 1. The flat directions considered}\end{center} 
\begin{center}
\begin{tabular}{|c|c|c|}          \hline
direction & dimension & ${\rm mass}^2$  \\ \hline
$H_uL$& $4$ &  $\frac 12 (m_H^2+ \mu^2_H+  m_{\tilde L_i}^2)$
 \\
$uude$& $4$ &  $\frac 14 ( m_{\tilde u_i}^2+ m_{\tilde u_j}^2+ m_{\tilde d_k}^2+
 m_{\tilde e_l}^2);~(i\ne j)$
 \\
$QQQL$& $4$ & $\frac 14 ( m_{\tilde Q_i}^2+ m_{\tilde Q_j}^2+ m_{\tilde Q_k}^2+ m_{\tilde L_l}^2;~
(i\ne j~{\rm or}~k)$
\\
$(udd)^{2}$& $6$ &  $\frac13 ( m_{\tilde u_i}^2+ m_{\tilde d_j}^2+ m_{\tilde d_k}^2);~(j\ne k)$
 \\
$(QLd)^{2}$& $6$ & $\frac13 ( m_{\tilde u_i}^2+ m_{\tilde d_j}^2+ m_{\tilde Q_k}^2)$
\\
\hline
\end{tabular}
\end{center}

\begin{figure}
\leavevmode
\centering
\vspace*{110mm}
\includegraphics{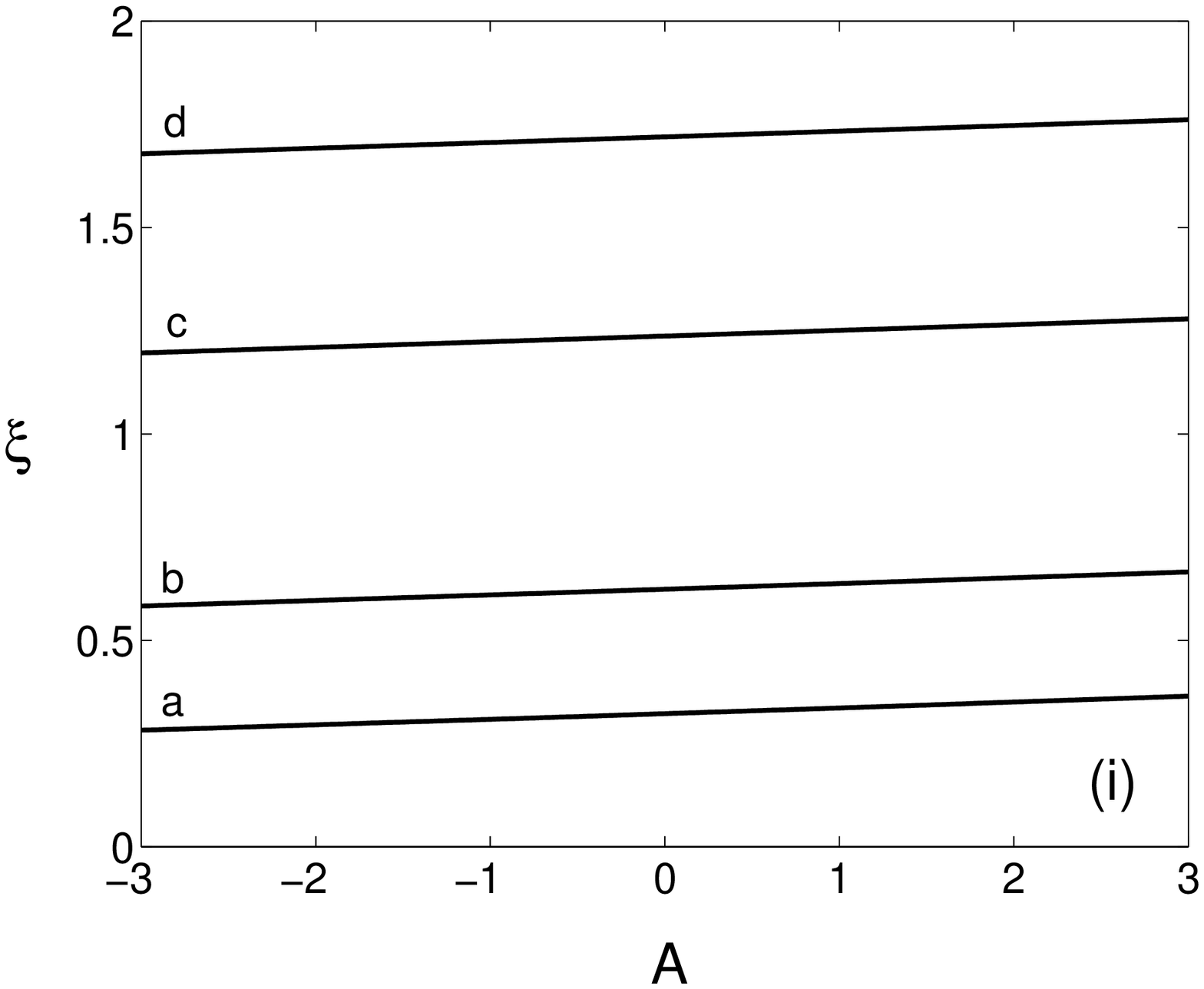}
\includegraphics{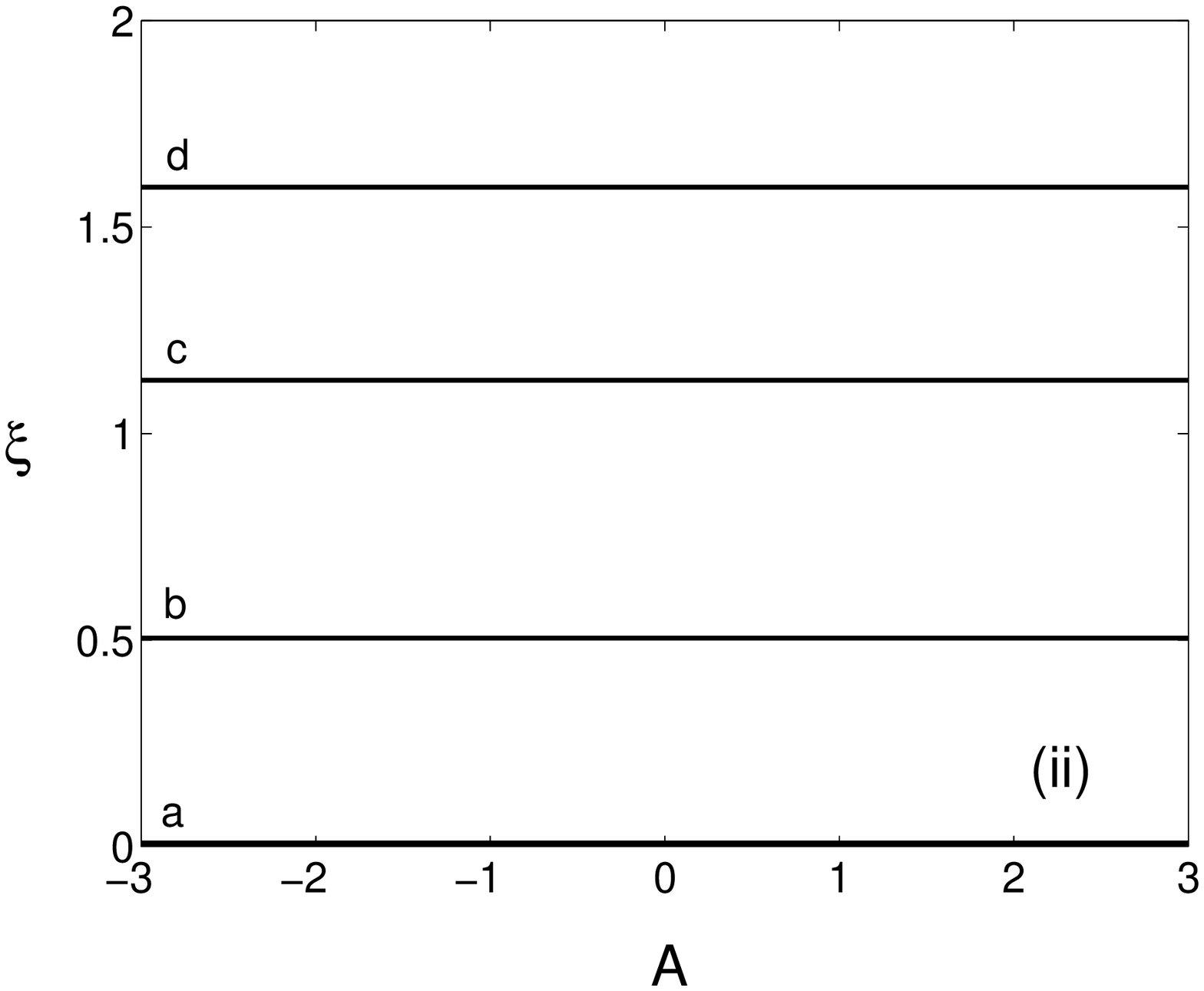}
\includegraphics{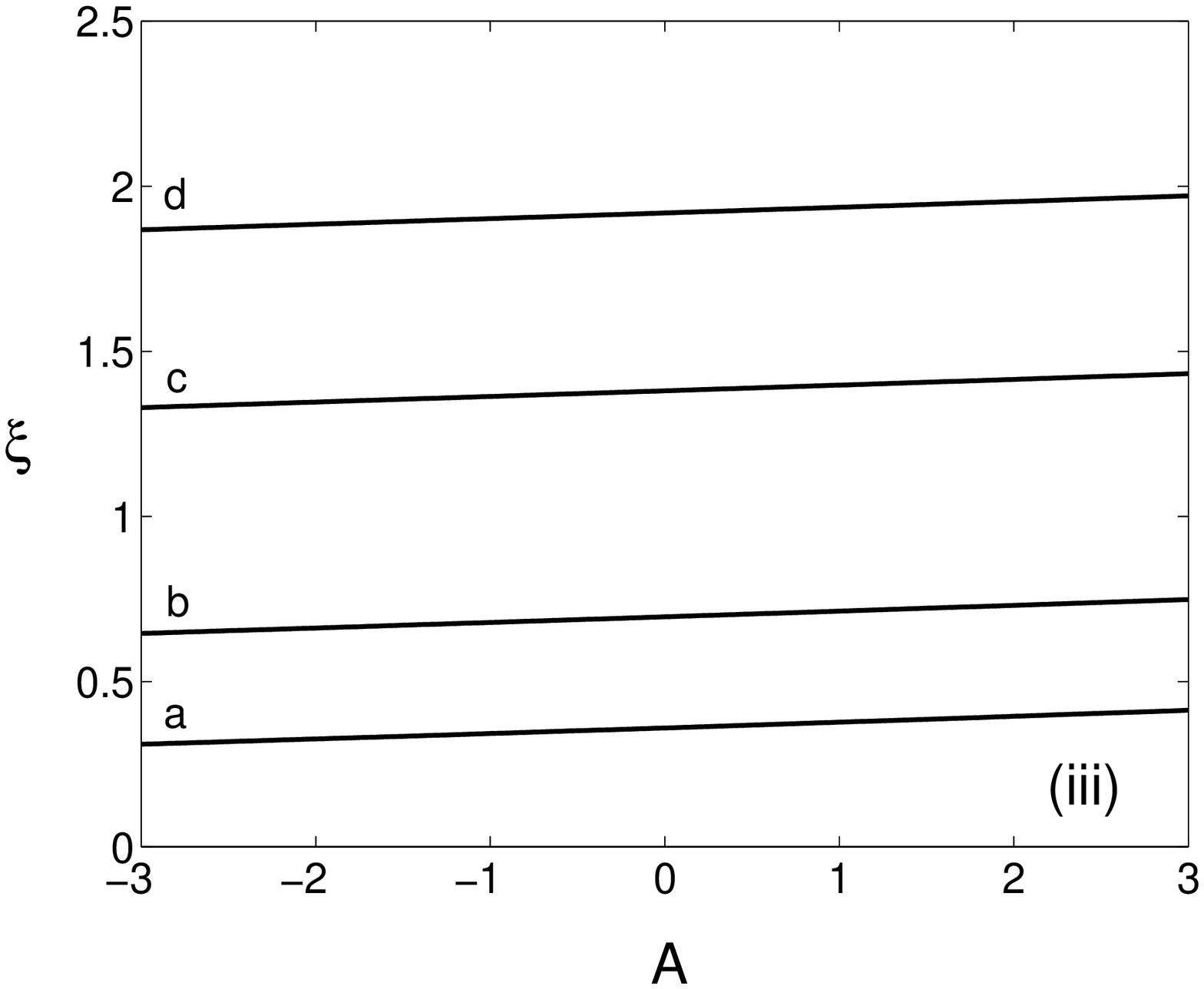}   
\includegraphics{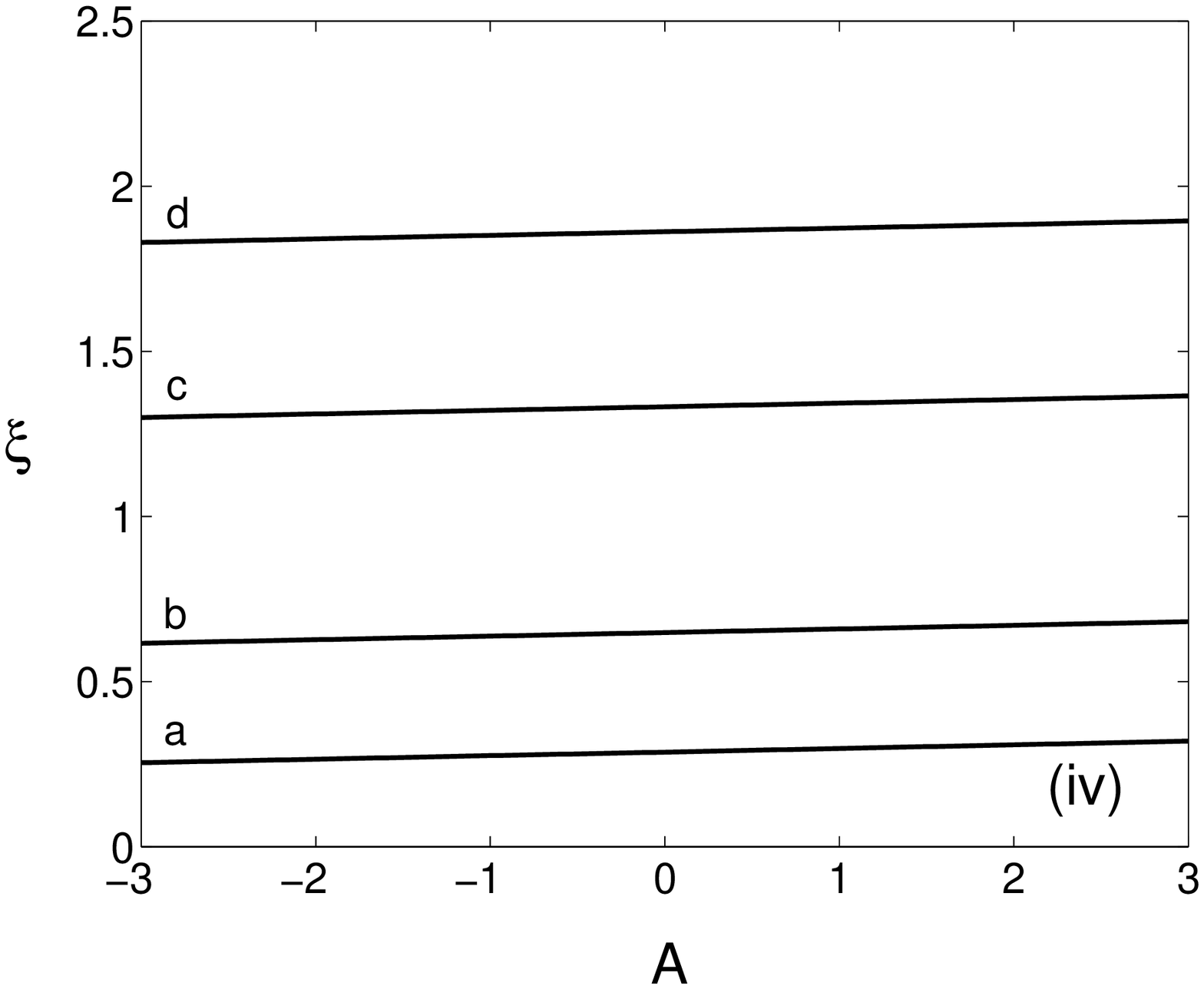}   
\caption{Contours of K for two d=4 flat directions: (a) $K=0$;
(b) $K=-0.01$; (c) $K=-0.05$; (d) $K=-0.1$. The
directions are (i) $Q_3Q_3QL$; (ii) $QQQL$, no stop; (iii) $u_3ude$; (iv) $uude$
with equal weight for all $u$-squarks.}
\label{kuva1}
\end{figure}
\begin{figure}
\leavevmode
\centering
\vspace*{100mm}
\includegraphics{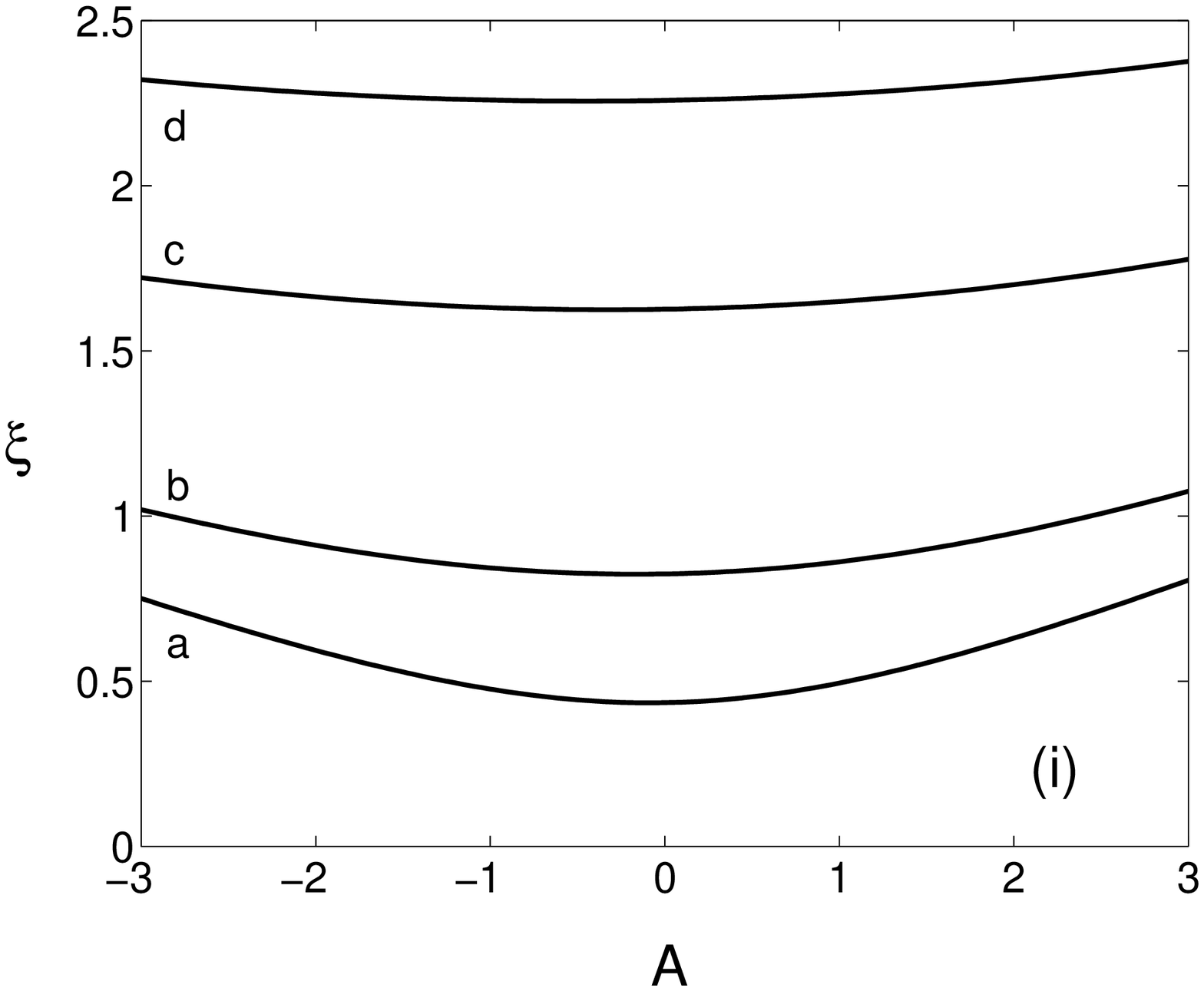}
\includegraphics{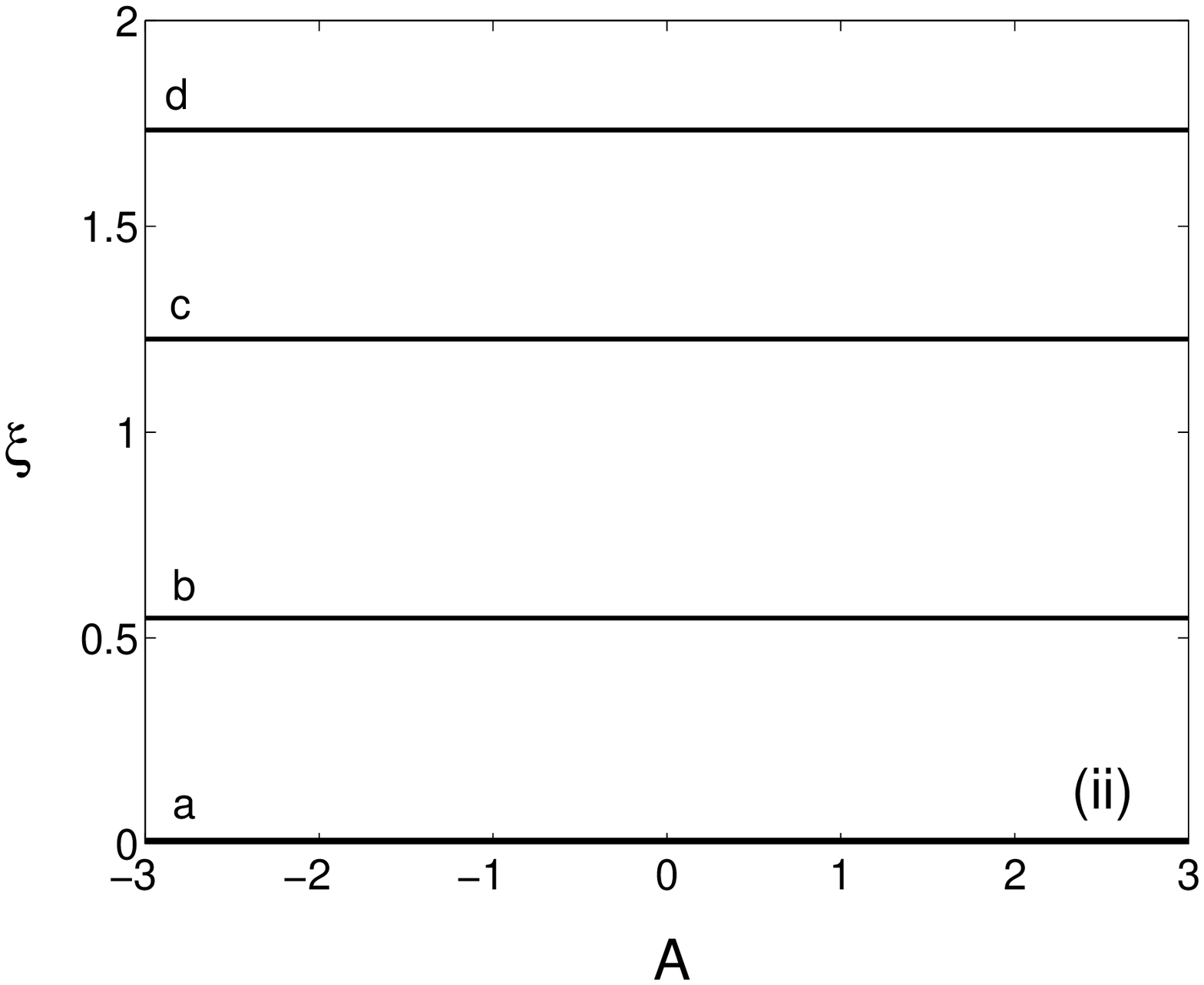}
\includegraphics{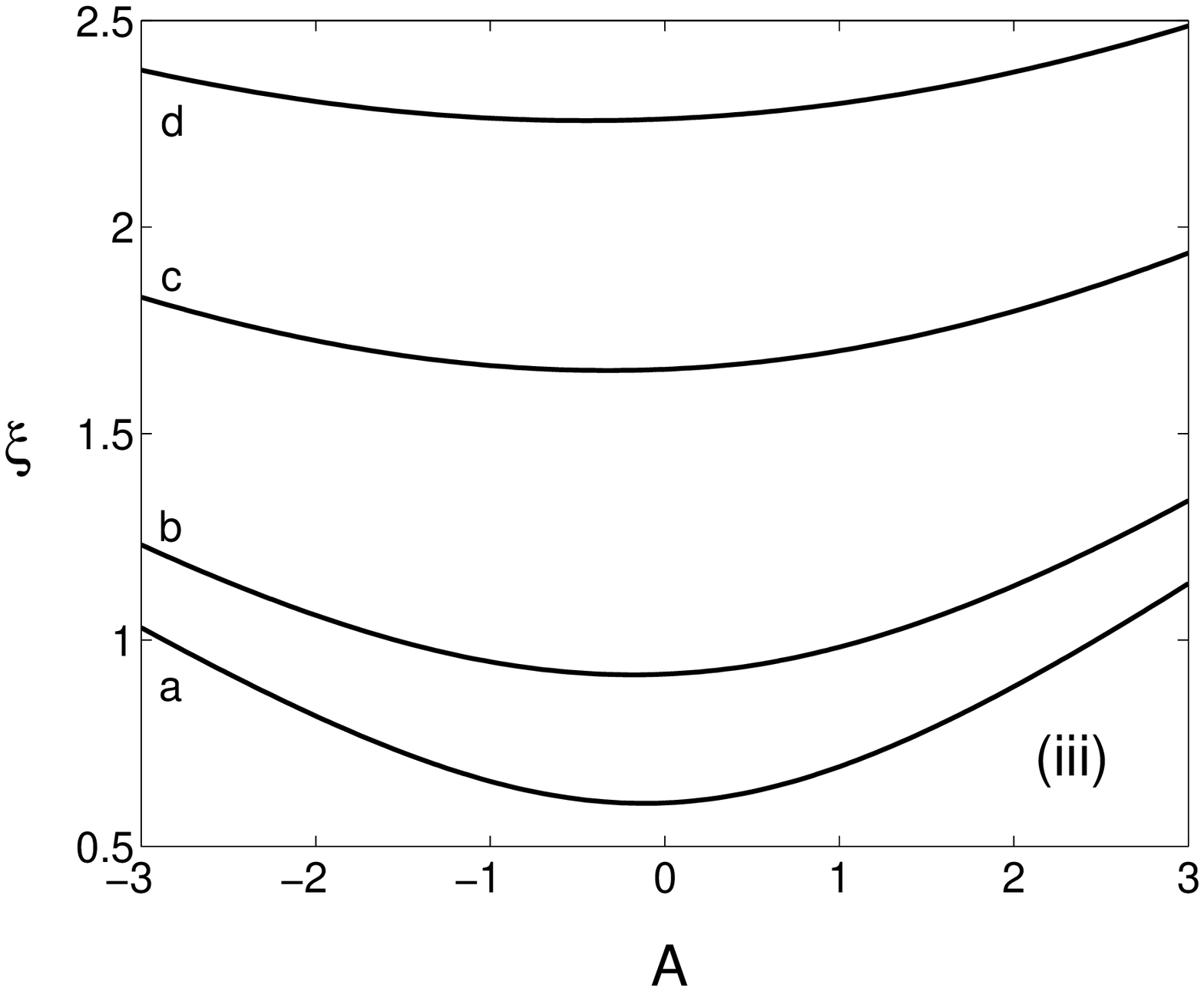}   
\includegraphics{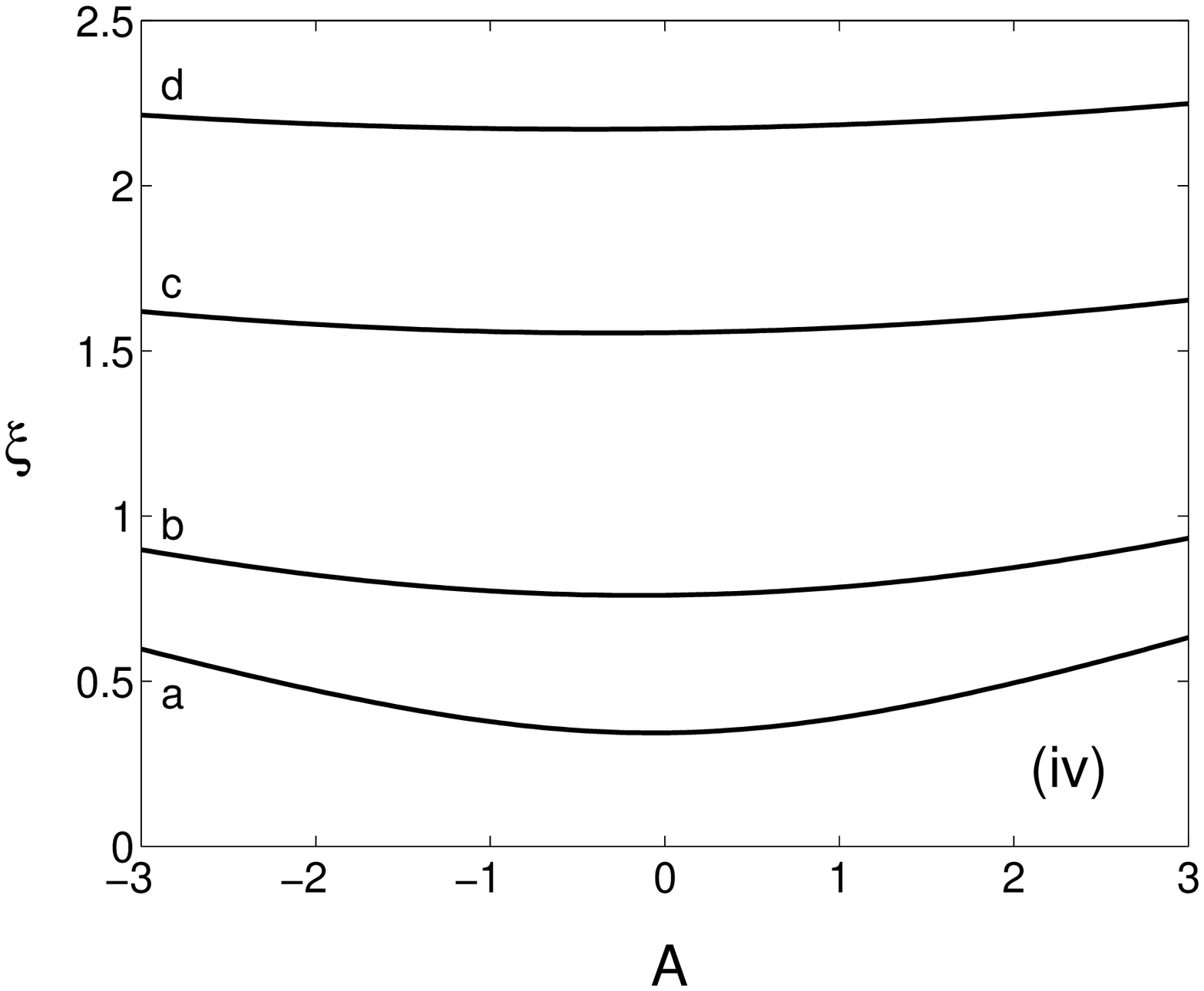}   
\caption{Contours of K for two d=6 flat directions: (a) $K=0$;
(b) $K=-0.01$; (c) $K=-0.05$; (d) $K=-0.1$. The
directions are (i) $Q_3Ld$; (ii) $QLd$, no stop; (iii) $u_3dd$; (iv) $udd$
with equal weight for all $u$-squarks.}
\label{kuva2}
\end{figure}
The flat directions we have studied are listed in Table 1. 
There is one purely leptonic direction, $H_uL$, a purely baryonic
one involving only the squarks, and directions which have both
squark and lepton fields.

In Fig. 1 
we show the contours
of $K$ for the d=4 $uude$ and $QQQL$ directions in the $(A, \xi)$ -plane;
Fig. 2 is for the d=6 $(udd)^{2}$ and $(QLd)^{2}$ directions.
These should be representative of all the other 
directions, too, except for $H_uL$. For $\xi\sim {\cal O}(1)$,
typical value for $K$ is found to be about $-0.05$.
For all the squark directions with no stop, as long as $h_b$ and
$h_u$ can be neglected,
$K$ is always negative, and the contours of equal $K$ do not depend
on $A$.
This is evident from the RGEs \eq{rge}. However, 
as far as flat directions are concerned, all squarks are
equal, and having no stop mixture would appear rather unnatural. Therefore 
we have considered the effect of stop mixing in the squark directions, 
which results in $A$-dependence of the $K$-contours, as is depicted in
Figs. 1 and 2. In the presence of stop mixing
$K<0$ is no longer automatic even in the purely squark directions.
 The more there is stop, the larger
value of $\xi$ is required for $K<0$. However, even for pure stop directions,
positive $K$ is typically obtained only for relatively light gaugino masses 
with $\xi\lsim 0.5$. In d=4 directions the effect of stop mixture is less 
pronounced than in the d=6 directions, as can be seen from Figs. 1 and 2.

In contrast to the squark directions,  $K$ was found to be always positive
in the $H_uL$-direction. This is due to the fact $H_uL$ does
not involve strong interactions which in other directions are mainly 
responsible for the decrease of the running scalar masses.
The value of $\mu_H$ was chosen in such a way that
for each value of $A$ and $\xi$, electroweak symmetry breaking is
obtained at the scale $M_W$.

Except for the $H_uL$-direction, the instability of the 
AD condensate is thus seen to depend on the amount of stop mixture
in the flat direction. Since it would be natural to expect roughly equal 
mixtures of the different squark flavours in the flat direction scalar, 
this means that in principle instabilities,
and hence Q-balls in the case of gravity mediated susy breaking,
could be ruled out experimentally at LHC by measuring the mass of
the gluino and the stop. To illustrate this, in Fig. 3 we show the
the domains of positive and negative $K$ in the region of 
$(m_{\tilde t},m_{\tilde g})$ -plane corresponding to the
range $-3<A<3$ and $0<\xi<2.5$. The Figure is for the $(udd)^{2}$ direction
with equal weight for all $u$-squarks. For most part $K>0$ and
$K<0$ regions can be separated, although there is a small area
below the upper $K=0$ contour where both values can be found. 
For a fixed $m$, the $K=0$ contour has endpoints which correspond
to $A=\pm 3$; if one were to allow for a wider range in $A$, this would
spread the region between the dashed lines towards the lower right-handed
corner. Changing the value of $m$ would redefine the physical mass
scale by a factor $m/100\GeV$. Thus measuring $m_{\tilde t}$ and 
$m_{\tilde g}$ would not alone be sufficient to  determine
the existence of instabilities. In addition, one needs the values of
$m$ and $tan\beta$ which naturally will be measured if 
supersymmetry will be found. Very roughly, instability is found
when $m_{\tilde g}\gsim m_{\tilde t}$, although 
the exact condition should be checked
case by case.

\begin{figure}
\leavevmode
\centering
\vspace*{90mm}
\includegraphics{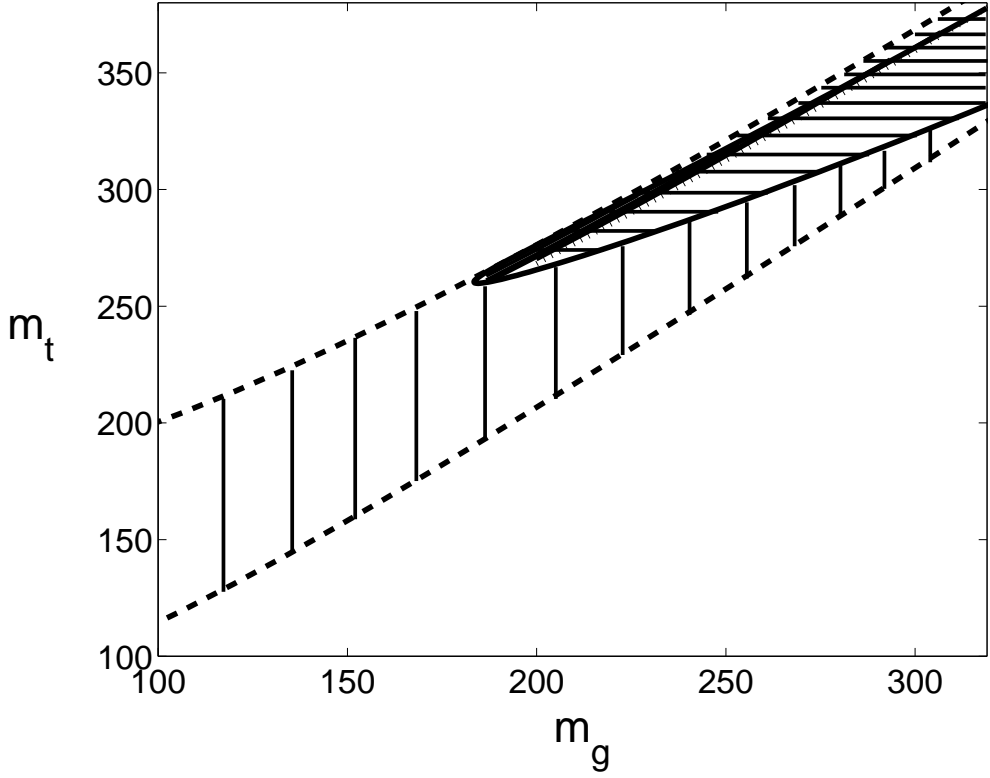}
\caption{The regions of positive and negative $K$ in the ($\tilde m_t,
m_{\tilde g}$)-plane in the interval $-3<A<3$ and $0<\xi<2.5$
(between the dashed lines) for the $udd$-direction with 
equal weight for all $u$-squarks.
 The solid line is the $K=0$ 
contour, horizontally hatched area is where $K<0$, vertically hatched area
is where $K>0$, and there is a small region below the upper
$K=0$ -line where both $K>0$ and $K<0$ can be found.
 The units are $(m/100\GeV)\GeV$.}
\label{kuva3}
\end{figure}

In conclusion, we have shown that the MSSM scalar condensates in 
all but the $H_uL$ flat direction are unstable for a large part
of the parameter space. Therefore the existence of Q-balls is
a generic feature in all the models that incorporate both the MSSM
and inflation. Moreover, as the existence of instabilities can in 
principle be ruled out by measuring the mass of the stop and the
gluino, one may soon be able to subject the Affleck-Dine scenario  
to a direct
test.

\subsection*{Acknowledgements}   This work has been supported by the
 Academy of Finland under the contract 101-35224 and the PPARC (UK). 

\end{document}